\documentclass[fleqn,usenatbib]{mnras}
\DeclareSymbolFont{cmletters}{OML}{cmm}{m}{it}
\DeclareMathSymbol{v}{\mathalpha}{cmletters}{"76}

\usepackage{ae,aecompl}
\usepackage{times}
\usepackage{soul}

\usepackage{tabularx,ragged2e,booktabs,caption}
\usepackage{amsmath}
\usepackage{amssymb}
\usepackage{epsfig}
\usepackage{graphicx}
\usepackage{ifthen}
\usepackage{latexsym}
\usepackage{rotating}
\usepackage{times,epsf}
\usepackage{txfonts}
\usepackage{varioref}
\usepackage{verbatim}
\usepackage{array}
\usepackage{url}
\usepackage{color}
\usepackage[T1]{fontenc}
\usepackage{epstopdf}
\usepackage{ulem}

\def\gsim{\mathrel{\raise.5ex\hbox{$>$}\mkern-14mu
             \lower0.6ex\hbox{$\,\sim$}}}
\def\lsim{\mathrel{\raise.3ex\hbox{$<$}\mkern-14mu
             \lower0.6ex\hbox{$\,\sim$}}}

\newcommand{\be}{\begin{equation}}
\newcommand{\ee}{\end{equation}}
\newcommand{\bea}{\begin{eqnarray}}
\newcommand{\eea}{\end{eqnarray}}

\title{Current closure in the pulsar magnetosphere}
\author[I. Contopoulos]
       {I. Contopoulos$^{1,2}$\thanks{E-mail: icontop@academyofathens.gr}\\
$^1$ Research Center for Astronomy and Applied Mathematics, Academy of Athens, Athens 11527, Greece\\
$^2$ National Research Nuclear University (MEPhI), Moscow 115409, Russia
}

\begin{document}

\maketitle

\label{firstpage}

\begin{abstract}
Current closure in the pulsar magnetosphere holds the key to its structure. We must determine not only the global electric circuit, but also the source of its electric charge carriers. We address this issue with the minimum number of assumptions: a) The magnetosphere is everywhere ideal and force-free, {\it except} above the polar cap and in some finite part of the current sheet; and b) pairs are produced above the polar cap with multiplicity $\kappa$. We show that a thin region of width $\delta\approx r_{\rm pc}/2\kappa\ll r_{\rm pc}$ along the rim of the polar cap provides all the charges that are needed in the equatorial and separatrix electric current sheet. These charges are transferred to the current sheet in a narrow dissipation zone just outside the magnetospheric Y-point. The maximum accelerating potential in this region is equal to the potential drop in the thin polar cap rim, which is approximately equal to $1/\kappa$~times the potential drop from the center to the edge of the polar cap. The dissipated electromagnetic energy is approximately equal to $0.5/\kappa$~times the total pulsar spindown energy loss. Our framework allows to calculate the high energy emission  in terms of the pair multiplicity.
\end{abstract}


\section{The magnetospheric electric current}
Live pulsars are powerful electromagnetic systems that transport Poynting energy from a central engine (the rotating magnetized neutron star), through a large scale electric circuit (the infinitely conducting magnetosphere), to electromagnetic loads at large distances (the termination shock and finite magnetospheric dissipative regions). The central `battery' is the electrically polarized conducting surface of the rotating magnetized neutron star, and the `infinitely conducting wires' lie along magnetic flux surfaces.
%\footnote{The force free condition $\rho_e {\bf E}+{\bf J}\times{\bf B}=0$ guarantees that electric currents flow along equipotential surfaces that coincide with magnetic flux surfaces.}.
Obviously, rotation polarizes not only the conducting surface of the central star, but also the conducting magnetosphere. Thus, the whole magnetosphere is filled with a distributed electric charge

During the past fifty years since we started painting the first theoretical picture of pulsars \citep{GJ69}, the astrophysical community has been mostly preoccupied with the origin and distribution of the magnetospheric electric charge which `is there' in order to guarantee that ${\bf E}\cdot{\bf B}=0$ everywhere in the magnetosphere. In particular, near the magnetic axis on the stellar surface, the electric charge density is given by the following approximate expression first derived by Goldreich and Julian, 
\begin{equation}\label{GJ}
\rho_{\rm GJ}\equiv \frac{\nabla\cdot{\bf E}_{*}}{4\pi}\approx -\frac{{\bf \Omega}\cdot{\bf B}_{*}}{2\pi c}\ .
\end{equation}
Here, $E_*$ and $B_*$ are the electric and magnetic fields around the stellar magnetic axis respectively. We will henceforth consider only the case of the aligned rotator in which the region above the polar cap is filled with a negative electric charge. In that case,
\begin{equation}
E_*=\frac{R}{r_{\rm lc}}B_*\ ,
\end{equation}
where, $R$ is the distance from the axis of rotation, and $r_{\rm lc}\equiv c/\Omega$ is the light cylinder radius. The counteraligned case is similar. The general 3D case will be considered in a future publication.

Soon after the original paper of Goldreich and Julian, it was proposed that the origin of the distributed magnetospheric electric charge are pair formation cascades above the magnetic poles. In these cascades, for every electron in the space charge above the stellar surface, $\kappa$ times more electron-positron pairs are produced that flow away from the magnetic poles at relativistic speeds \citep{DH92}. $\kappa$ may be as high as $10^3$ or even $10^5$ for young pulsars \citep[][and references therein]{TH15}, but decreases to values of order unity for charge starved magnetospheres off old pulsars, and drops down to zero in dead pulsars.

The astrophysical community hasn't paid equal attention to the origin and distribution of the magnetospheric electric current. In a conducting magnetosphere, the electric current is the agent that generates spindown torques onto the stellar surface, and thus carries stellar rotational energy to large distances in the form of electromagnetic (Poynting) flux. One naive estimate would be that the electric current density just above the polar cap is on the order of $\rho_{\rm GJ}c$. This would correspond to the space charge flowing out everywhere at the speed of light. \citet{CKF99} (hereafter CKF) tought us that this is not the case. In fact, the electric current density along magnetic field lines that originate in the so-called polar cap differs significantly from the naive estimate, and even has the opposite sign in certain parts of the polar cap \citep[e.g.][]{PS18}. The total electric current $I_{\rm pc}$ through the polar cap, though, may still be estimated roughly as
\begin{equation}\label{Iestimate}
I_{\rm pc}\sim \rho_{\rm GJ}c\pi r_{\rm pc}^2 \approx \frac{\Omega\Psi_{\rm pc}}{2\pi}\ ,
\end{equation}
where, $\Psi_{\rm pc}$ is the total amount of magnetic flux contained in the polar cap, $r_{\rm pc}\approx (r_*^3/r_{\rm lc})^{1/2}$ is the radius of the polar cap (approximately centered on the magnetic axis), and $r_*=10$~km is the radius of the neutron star. Notice that our expressions for the polar cap radius and electric current are only approximate, i.e. they are missing factors of order unity obtained in the CKF solution. In the present work, we define as `polar cap' the area around the magnetic axis from which originate magnetic field lines that cross the light cylinder. Obviously, once the electromagnetic (Poynting) flux carried by these field lines crosses the light cylinder, it cannot return back to the star. Eq.~(\ref{Iestimate}) yields a rough estimate of the electromagnetic losses of the neutron star as \begin{equation}
\dot{E}\sim 2I_{\rm pc}V_{\rm pc}\approx \frac{\Omega^2\Psi_{\rm pc}^2}{2\pi^2 c}\approx 3\times 10^{33}\ \left(\frac{B_*}{10^{13}\ {\rm G}}\right)^{2}\left(\frac{P}{1\ {\rm s}}\right)^{-4}\ {\rm erg}\ {\rm s}^{-1}\ ,
\label{4}
\end{equation}
where, $V_{\rm pc}=\int_0^{r_{\rm pc}}E_* {\rm d}R=r_{\rm pc}^2B_*/2r_{\rm lc}=\Psi_{\rm pc}/2\pi r_{\rm lc}$ is the potential drop from the center to the periphery of the polar cap. The factor of two in eq.~(\ref{4}) is due to the contribution from both polar caps.

The electric circuit originating in the polar cap must somehow close. CKF first showed that magnetospheric electric current closure in the interior of the polar cap is only partial. {\it The bulk of the current closes in the form of a current sheet in a `separatrix' region between magnetic field lines that cross the light cylinder, and magnetic field lines that close back onto the star}. This is a singular region in which the ideal relativistic force-free MHD notion that `currents flow along magnetic flux surfaces' does not apply. To be more precise, the electric current circuit closes along and outside (not inside!) the boundary of the polar cap. The separatrix region has nothing to do with the so-called slot gap which originates entirely inside the polar cap \citep{MH03}.

\section{Charge carriers in the return current}

The density of charge carriers in a current sheet is practically `infinite'. In other words, the electric charge density along the separatrix is much greater than and has nothing to do with the classical Goldreich-Julian estimate of eq.~(\ref{GJ}). The current sheet itself has a surface charge density $\sigma$ that we can easily estimate near the surface of the star as follows: Interior to the return current sheet, the magnetic field consists of a dipolar magnetic field component $B_*$ and an extra azimuthal magnetic field component $B_\phi= 2I_{\rm pc}/r_{\rm pc}c$ due to the magnetospheric electric current that threads the polar cap. Outside the current sheet, the field consists only of a dipolar magnetic field component $B_{*{\rm OUT}}$. Continuity of $B^2-E^2$ accross the separatrix region implies a discontinuity of $B_*$, thus also a discontinuity of $E_*$ equal to
\begin{equation}\label{dE}
E_{{\rm OUT}\ *}-E_{{\rm IN}\ *}\approx -\frac{1}{2}\left(\frac{r_{\rm pc}}{r_{\rm lc}}\right)^3 B_*\equiv 4\pi\sigma_{\rm pc}<0\ .
\end{equation}
If we assume that this surface charge moves {\it inwards} at the speed of light, it generates a separatrix electric current equal to $2\pi r_{\rm pc} c \sigma_{\rm pc}\sim (r_{\rm pc}/r_{\rm lc})^2 I_{\rm pc}/2$ which is in absolute terms much smaller than $I_{\rm pc}$. Obviously, the separatrix charge at the edge of the polar cap is not sufficient to support the separatrix return electric current. An important question thus arises: what is the origin of the charge carriers in the separatrix return current?

This question has been addressed only very recently. \citet{CKK14} discussed the limit in which the separatrix return current vanishes. In that limit, the total magnetospheric current closes through the polar cap, but it can only do so via a dissipation zone in the equatorial current sheet beyond the light cylinder. This solution yields about $40\%$ dissipation of the total spindown energy in that region. This solution requires a space charge separated magnetosphere in which the positrons required to provide the charge and the electric current in the equatorial current sheet are provided by the magnetic field lines that enter it all along. There are no extra pairs to worry about in the current closure, and the magnetic field lines that open up to infinity are negatively charged and carry the outgoing electric current.

More recently, \citet{Betal18} addressed the same issue focusing on the (numerically imposed) pair formation multiplicity in the polar cap. They obtained various solutions in which particles that originate near the boundary of the polar cap gradually enter the separatrix current sheet all along its length. Similar results where also obtained by \citet{CPPS15}. The problem with these works is that they are asking too much from global (so-called `ab initio') PIC simulations. The current state-of-the-art PIC simulations follow unrealistically heavy super-particles that flow through the magnetosphere like `billiard balls' \citep{C16}. This can be seen directly in figures~(11)-(13) of \citet{Betal18} in which the trajectories of the heavy particles are clearly unrealistic. In another work of the same group \citep{Ketal18}, there is an interesting discussion on the limitations of PIC simulations in capturing particle trajectories in the asymptotic drift approximation \citep[also called `Aristotelian';][]{G13}.

\begin{figure}
 \centering
 \includegraphics[width=7cm,height=5cm,angle=0.0]{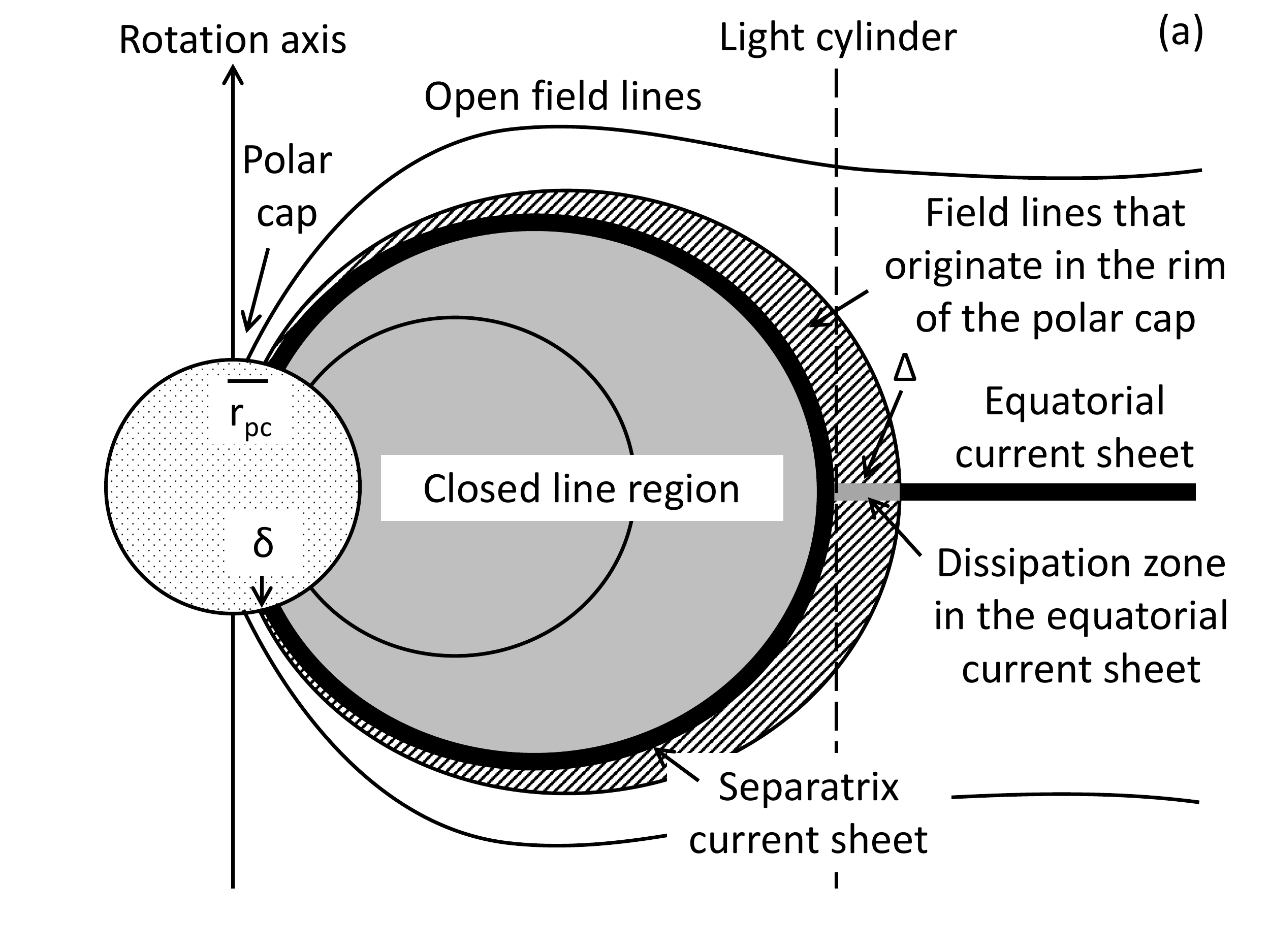}
 \includegraphics[width=7cm,height=5cm,angle=0.0]{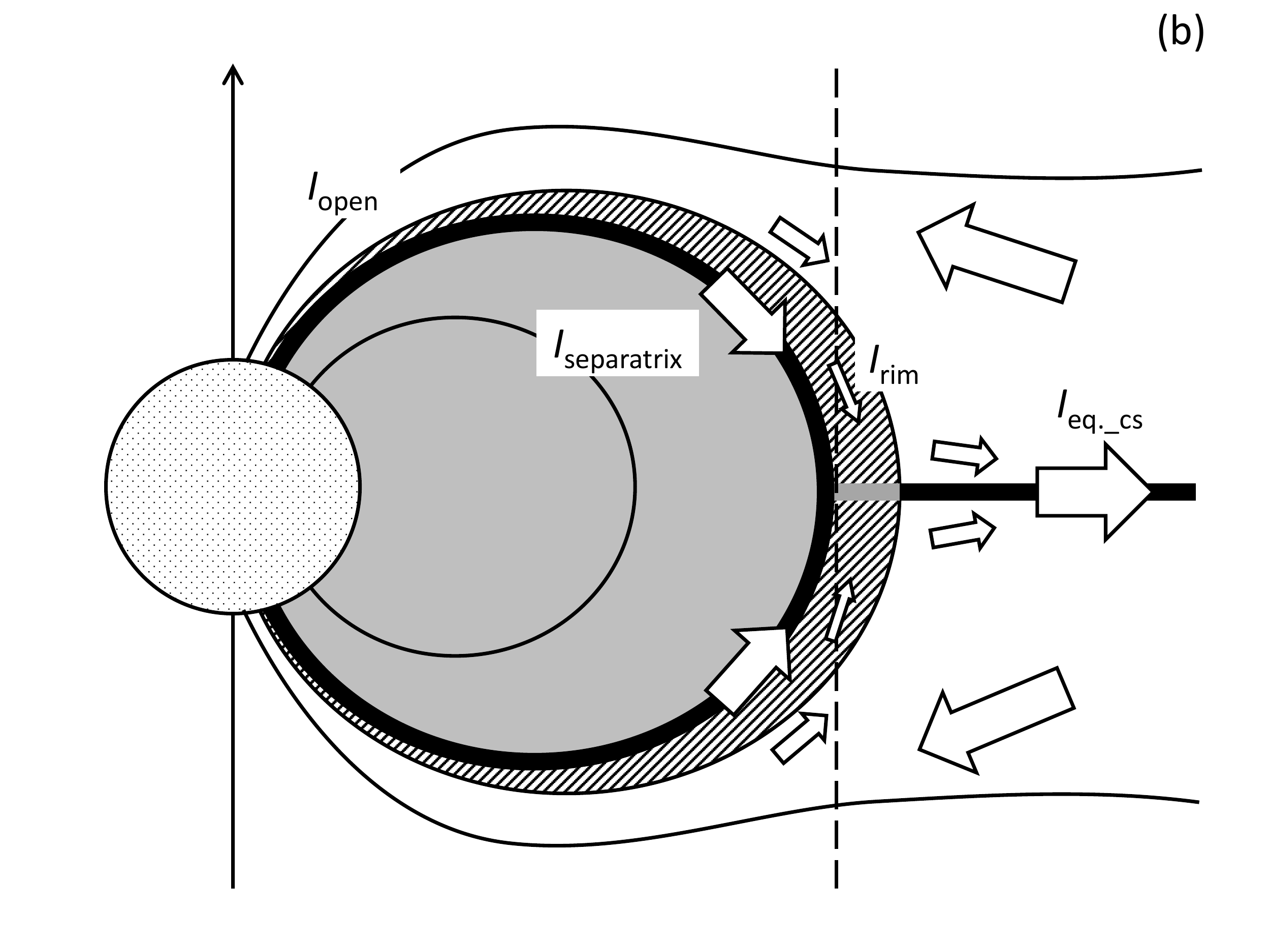}
 \includegraphics[width=7cm,height=5cm,angle=0.0]{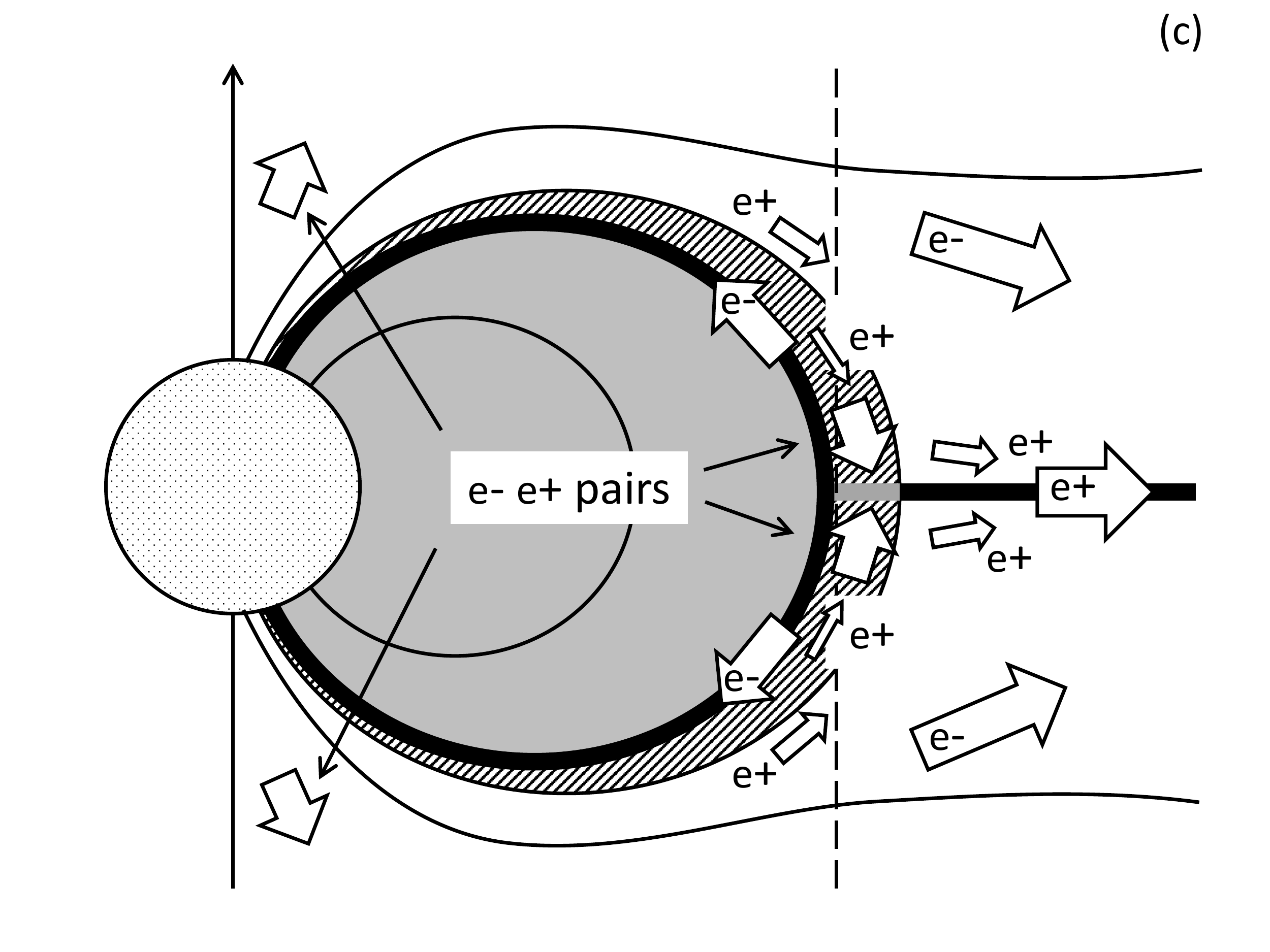}
\caption{Current closure in the pulsar magnetosphere (not drawn to scale: $\delta\ll r_{\rm pc}\ll r_*\ll r_{\rm lc}$, $\delta\ll \Delta \ll r_{\rm lc}$.). Central sphere: neutron star. Dashed line: light cylinder. Grey region: closed field lines. White region: open field lines. Striped region: field lines that originate in the rim of the polar cap and close in an equatorial dissipation zone (thick grey line) just outside the light cylinder. Thick black line: Return current sheet on the surface of and outside the striped region; (b) Current circuit: $I_{\rm pc}=I_{\rm open}-I_{\rm rim}=I_{\rm separatrix}$, $I_{\rm open}=I_{\rm eq.\_cs}/2$; (c) Charge carriers in the electric current circuit. Pairs are generated only above the polar cap.}
\label{figure1}
\end{figure}

We here propose a simple solution to the question of the origin of the charge carriers in the return current. We will only consider for simplicity the limit of very large pair multiplicity $\kappa\gg 1$ in the axisymmetric regime. The general case of small and large multiplicites in a 3D misaligned rotator will be discussed in a future numerical work. Let us begin by considering a narrow ring of width $\delta\ll r_{\rm pc}$ along the rim of the polar cap on the stellar surface. This carries an amount of magnetic flux equal to
\begin{equation}
\Psi_{\rm rim}=2\pi r_{\rm pc}\delta B_* = \frac{2\delta}{r_{\rm pc}}\Psi_{\rm pc}\ .
\end{equation}
According to several numerical solutions in the literature, this region also carries a small amount of return current $I_{\rm rim}$ \citep[e.g.][]{CKF99, T06}. It also carries a flux of electron-positron pairs equal to $2\pi r_{\rm pc}\delta \kappa \rho_{\rm GJ} c/e$. The electron content in that narrow region is sufficient to support the return current $I_{\rm separtrix}$ in the separatrix, 
provided that
\begin{equation}\label{delta}
\delta\approx\frac{I_{\rm separatrix}}{\kappa r_{\rm pc}\Omega B_*}\ .
\end{equation}
The natural way for these pairs to supply the electrons needed to form the separatrix return current is the {\it development of a dissipation zone just outside the Y-point at the origin of the equatorial current sheet on the light cylinder} (see description in Figure~1). In that region, the pair electrons will be accelerated inwards towards the light cylinder, and will then flow back to the star along the separatrix, whereas the pair positrons and the extra positrons that support the positive electric charge and the return current $I_{\rm rim}$ in that region will flow outwards. The combined outward flow of positrons will supply the total return electric current and the electric charge in a dissipationless equatorial current sheet beyond the narrow dissipation zone. The maximum energy that may be attained by the electrons and positrons that enter the dissipation zone is
\begin{equation}\label{Eacc}
\epsilon_{\rm max} = eV_{\rm rim} = \frac{2\delta}{r_{\rm pc}}eV_{\rm pc}\ ,
\end{equation}
where, $V_{\rm rim}\ll V_{\rm pc}$ is the potential drop accross the rim of the polar cap. As we will see in the next section, in young pulsars, radiation reaction will limit the maximum particle energy to lower values.

When the pair formation multiplicity is high, the magnetospheric structure is very-very close to the ideal CKF solution, with only a small amount of dissipation outside the light cylinder \citep{KHK14, PS14, Ketal18}. In that case, $I_{\rm rim}\ll I_{\rm pc}$, $I_{\rm separatrix}\approx I_{\rm pc}$, and therefore, eqs.~(\ref{delta}), (\ref{Iestimate}) and (\ref{Eacc}) yield
\begin{equation}
\delta\approx\frac{r_{\rm pc}}{2\kappa}\ll r_{\rm pc}\ ,\ \epsilon_{\rm max}\approx \frac{eV_{\rm pc}}{\kappa}\ll eV_{\rm pc}\ .
\end{equation}
What is very important is that our model also allows us to estimate the rate of energy dissipation into high-energy luminosity $L_\gamma$ in the pulsar magnetosphere. This is equal to the Poynting flux carried by the magnetic field lines that originate in the polar cap rim. A straightforward calculation yields
\begin{equation}
L_\gamma\approx \frac{\delta}{r_{\rm pc}}\dot{E}=\frac{\dot{E}}{2\kappa}\ .
\end{equation}
This simple relation tells us that for typical $\gamma$-ray pulsars with observed $\gamma$-ray luminosity efficiencies equal to $0.1\%-1\%-10\%$, the corresponding pair formation multiplicities are expected to lie in the range $\kappa\approx 500-50-5$ respectively. Notice also that, if we assume that $L_\gamma/\dot{E}$ is roughly proportional to $\dot{E}^{-1/2}$ as suggested by observations \citep{T08, GH15}, then $\kappa\propto \dot{E}^{1/2}$.

A final comment is in order here before we move on to the next section. As we saw in the calculation below eq.~(\ref{dE}), near the surface of the star the separatrix electric charge fails to account for the separatrix return current by a factor $(r_{\rm pc}/r_{\rm lc})^2\ll 1$. The same calculation near the Y-point yields the separatrix surface charge density $\sigma_{\rm lc}$ near the light  
cylinder $R\rightarrow r_{\rm lc}$,
\begin{equation}
E_{{\rm OUT}\ {\rm lc}}-E_{{\rm IN}\ {\rm lc}}\approx 
-\frac{-2B_{\rm lc}}{[1-(R/r_{\rm lc})^2]^{1/2}}\equiv 4\pi \sigma_{\rm lc}<0
\end{equation}
\citep[see][for details]{U03, T06}. Here, $B_{\rm lc}\equiv B_*(r_*/r_{\rm lc})^3/2$ is the value of the dipole magnetic field at the light cylinder. In the above estimate, `IN' and `OUT' at the Y-point refers to the region just outside and inside the light cylinder respectively. Obviously, the separatrix electric charge is sufficient to support the separatrix return current. The only way to reconcile this  result with the inefficiency of the separatrix electric charge to carry the same electric current near the surface of the star is the following: The separatrix must contain {\it an extra amount of distributed corrotating positive electric charge} with surface charge density $\sigma_+(R)>0$ such that,
\begin{equation}
\sigma_+(R) + \sigma_-(R) = \sigma(R)
\end{equation}
at all distances, from the polar cap at $r_{\rm pc}$ to the light cylinder. Here, $\sigma_-(R)\sim -I_{\rm pc}/2\pi R c$ is the negative charge density carried by the electrons that inflow from the dissipation zone. Clearly,
\begin{equation}
\sigma_{-\ {\rm pc}}\sim -\frac{r_{\rm pc}}{r_{\rm lc}}\frac{B_*}{4\pi}=\left(\frac{r_{\rm lc}}{r_*}\right)^3\sigma_{\rm pc}\ ,
\end{equation}
which is in absolute terms much greater than $\sigma_{\rm pc}$. Consequently, $\sigma_{+\ {\rm pc}}\approx |\sigma_{-\ {\rm pc}}|$. 
It is conceivable that these corotating positrons are the result of pair formation by the inflowing separatrix electrons. 
Note that the concept of one particle population of the `right' type flowing through another stationary (corotating) population of the `wrong' type needed to support both the electric charge and electric current distributions in the pulsar magnetosphere was first proposed in section 4 of \citet{C16}.

\section{High energy radiation}

If we know the accelerating electric field and the trajectory of particles in the equatorial dissipation region, it is straightforward to calculate the asymptotic particle Lorentz factor $\Gamma$ in the curvature radiation reaction limit, 
\begin{equation}\label{radrl}
\frac{{\rm d}\Gamma}{{\rm d}t}=\frac{ecE_{\rm acc}}{m_ec^2}-\frac{2e^2\Gamma^4}{3r_{\rm lc}^2m_ec}=0\Rightarrow \Gamma_{\rm rrl}=\left(\frac{3r_{\rm lc}^2 E_{\rm acc}}{2e}\right)^{1/4}\ .
\end{equation}
Here, $E_{\rm acc}$ is the accelerating electric field in the dissipation region. We have assumed here that the radius of curvature around the Y-point is equal to $r_{\rm lc}$. We can estimate $E_{\rm acc}$ if we make the simplifying assumption that the magnetic flux that crosses the equatorial dissipation zone does so in a dipole-like configuration. In that case, the width $\Delta$ of that region is equal to
\begin{equation}
\Delta \approx 2\delta \left(\frac{r_{\rm lc}}{r_*}\right)^{3/2}
\approx \frac{r_{\rm lc}}{\kappa}\ ,
\end{equation}
and hence, the accelerating electric field in that region is equal to 
\begin{equation}
E_{\rm acc}\approx \frac{V_{\rm rim}}{\Delta}\approx B_{\rm lc}\ .
\end{equation}
We have thus shown that the accelerating electric field is approximately equal to the value of the magnetic field near the light cylinder. One can easily check that, with this accelerating electric field, 
\begin{equation}
\Gamma_{\rm rrl}=4\times 10^7\left(\frac{B_*}{10^{13}{\rm G}}\right)^{1/4}\left(\frac{P}{1\ {\rm s}}\right)^{-1/4}\ ,
\label{Gammarrl}
\end{equation}
where, $P$ is the pulsar period. This will be reached within an acceleration distance on the order of
\begin{equation}
l_{\rm acc} \equiv \frac{\Gamma_{\rm rrl}\ m_e c^2}{eE_{\rm acc}}\approx 0.4\ \left(\frac{B_*}{10^{13}\ {\rm G}}\right)^{-3/4}\left(\frac{P}{1\ {\rm s}}\right)^{7/4}r_{\rm lc}\ .
\end{equation}
Obviously, the acceleration in the equatorial dissipation region will be able to reach the radiation reaction limit only if $l_{\rm acc}\le \Delta$, i.e. if $P\lsim \kappa^{-4/7}$~s. This will be satisfied only for young pulsars. For slower pulsars, the radiation reaction limit is probably not reached in the equatorial dissipation region. In that case, the particles that will benefit from the full potential drop accross the rim region will be accelerated to a maximum Lorentz factor $\Gamma_{\rm max}$ equal to
\begin{equation}
\Gamma_{\rm max}=\frac{eE_{\rm acc} \Delta}{m_e c^2}=
\frac{eB_* r_*^3}{2\kappa r_{\rm lc}^2 m_e c^2}
=\frac{10^8}{\kappa}\left(\frac{B_*}{10^{13}\ {\rm G}}\right)\left(\frac{P}{1\ {\rm s}}\right)^{-2}\ .
\label{Gammamax}
\end{equation}
The curvature radiation $\gamma$-ray cutoff energy will thus be determined by the smallest of $\Gamma_{\rm rrl}$ and $\Gamma_{\rm max}$, namely
\begin{eqnarray}
\epsilon_{\rm cut} & = & \min(\Gamma_{\rm max}^3,\Gamma_{\rm rrl}^3)\ \frac{3hc}{2r_{\rm lc}}\nonumber\\
& = & \frac{80}{\kappa^3}\left(\frac{B_*}{10^{13}\ {\rm G}}\right)^3\left(\frac{P}{1\ {\rm s}}\right)^{-7}{\rm GeV}
\ \ \ {\rm if}\ \Gamma_{\rm max}<\Gamma_{\rm rrl}\ ,
\label{cutoff1}\\
&  & 3\left(\frac{B_*}{10^{13}\ {\rm G}}\right)^\frac{3}{4}\left(\frac{P}{1\ {\rm s}}\right)^{-\frac{7}{4}}\,{\rm GeV}
\ \ \ \ {\rm if}\ \Gamma_{\rm max}>\Gamma_{\rm rrl}\ .
\label{cutoff2}
\end{eqnarray}
We see that the spectrum is limited by radiation reaction only at small multiplicities. At large multiplicities, the spectrum cutoff energy decreases as $\kappa^{-3}$. Notice also that if $\kappa$ is roughly proportional to $\dot{E}^{1/2}$, then $\epsilon_{\rm cut}$ is proportional to $P^{-1}$ in the accelerating regime, and to $\dot{E}^{3/8}$ in the radiation reaction limit.

The detailed $\gamma$-ray spectrum will be a superposition of the spectra of the particles that enter the dissipation zone at various positions (see Figure~2).
Synchrotron emission beyond the Y-point may also contribute to the high-energy spectrum, and in some cases may even be more efficient than curvature radiation \citep[according to][]{PS18}.
%In the Appendix we present a first calculation of the resulting spectrum. 
Accurate estimates of spectra and 3D light curves require detailed PIC simulations of the equatorial dissipation region near and outside the Y-point. 
The work of Kalapotharakos and collaborators has shown that high-energy emission inside the light cylinder (e.g. in so-called ``outer gaps'')  does not easily result in light curves that are consistent with Fermi observations \citep[e.g.][]{KHK14}, thus our present model is promising.
Notice that we do not need to perform global (`ab initio') PIC simulations since the rest of the magnetosphere is practically ideal and force-free. We thus plan to focus our computational resources only in the vicinity of the Y-point.

\begin{figure}
 \centering
 \includegraphics[width=7cm,height=5cm,angle=0.0]{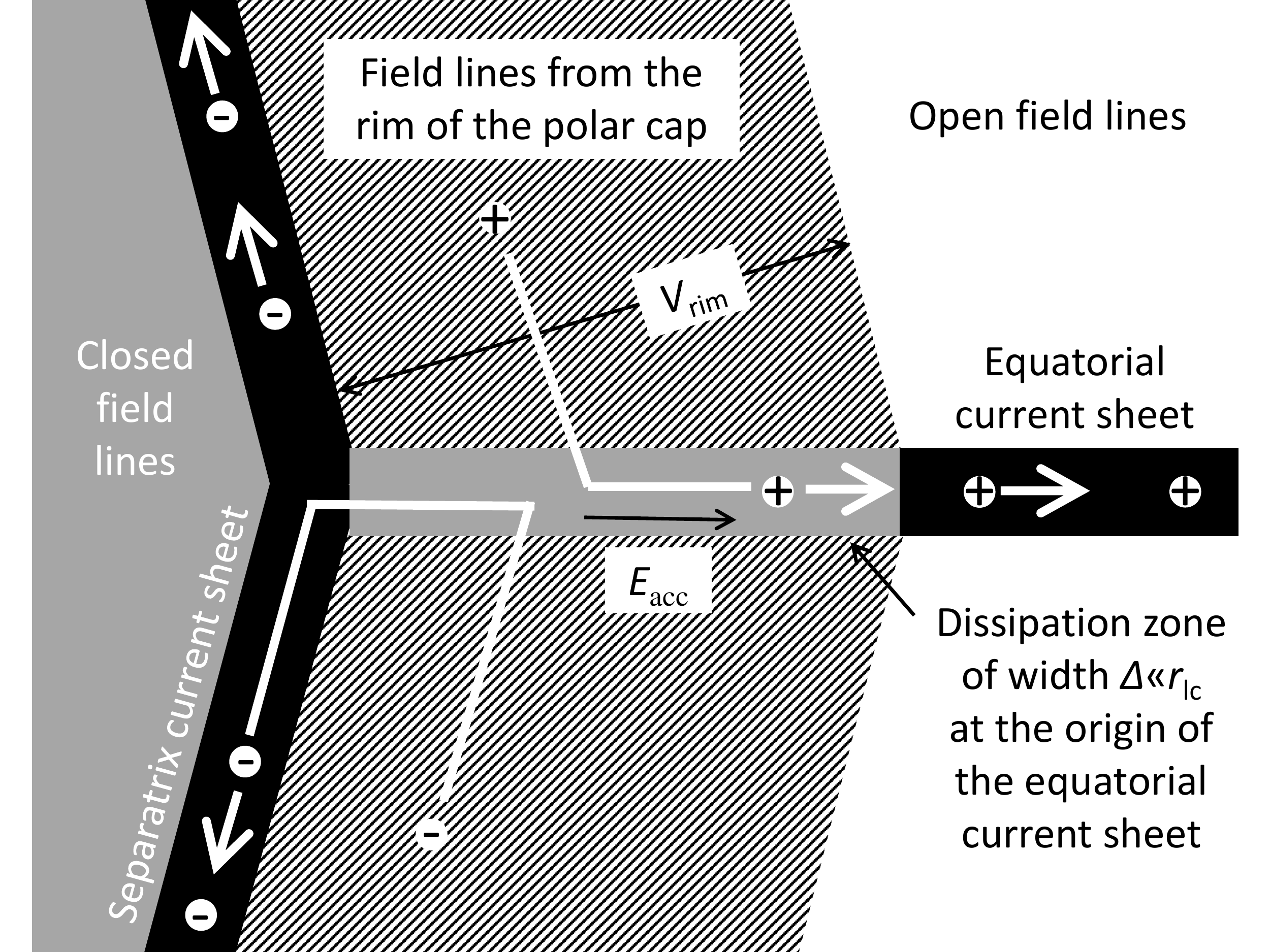}
\caption{Detail of the equatorial dissipation zone (see Figure~1a for details). While in that region, positrons that originate over the rim of the polar cap are accelerated outwards by the accelerating field $E_{\rm acc}$ and radiate curvature radiation. After they exit the dissipation region, they populate the equatorial current sheet. Similarly, electrons that too originate over the rim of the polar cap are accelerated and radiate inwards, and they populate the separatrix current sheet.}
\label{figure2}
\end{figure}

\section{Summary}

We have here proposed the following solution to the issue of the current closure in the pulsar magnetosphere:
\begin{enumerate}
\item The pulsar magnetosphere is everywhere ideal and force-free except for the region above the polar cap (where pairs are generated), the termination shock at `infinity', and an equatorial narrow dissipation zone at the origin of the equatorial current sheet just outside the magnetospheric Y-point. Magnetic field lines from the rim of the polar cap enter the latter region after crossing the light cylinder, and return to the other polar cap.
\item The electric charge carriers in the separatrix are provided by one sign of the pairs in the polar cap rim. It is also conceivable that these inflowing particles will also form pairs near the star and thus contribute to the general magnetospheric pair multiplicity.
\item The other sign of the electric charge carriers, together with the electric current that outflows from the polar cap rim, forms the rest of the return current in the equatorial current sheet beyond the dissipation zone.
\item The rest of the magnetic field lines in the polar cap opens up to infinity. Equivalently, the equatorial current sheeet beyond the dissipation zone is practically dissipationless.
\end{enumerate}

The physical motivation for the development of the dissipation zone at the particular position that we propose is clear: it acts as the source of the right type of charges in both the equatorial and separatrix current sheets. This can be described as ``pair generation without pair formation'' at the Y-point (the required pairs are carried to the dissipation zone from the polar cap rim). Without a source of pairs at the Y-point, all available positrons will move out, electrons will move in, thus, the region around the Y-point will be depleted of the needed charge carriers, ideal force-free conditions will brake down, and the dissipation zone will develop.

We expect that high-energy radiation is produced only wherever particles are accelerated, since otherwise their energies are quickly radiated away via curvature radiation. In the framework of our simple model, particles are accelerated only in the equatorial dissipation zone and in the polar caps, and therefore, high-energy radiation is produced {\it nowhere else} in the magnetosphere. This simplifies very much our efforts to understand how $\gamma$-ray pulsars operate. In particular, magnetospheric solutions that correspond to pair formation multiplicities $\kappa\gsim 10$ (or equivalently $\gamma$-ray efficiencies $L_\gamma/\dot{E}\lsim 10\%$) will differ only slightly from the standard ideal force-free CKF-type solution. In that case, we are able to determine the maximum particle energy (eqs.~\ref{Gammarrl}, \ref{Gammamax}) and the resulting curvature radiation cutoff energies (eqs.~\ref{cutoff1}, \ref{cutoff2}) as functions of the pulsar period $P$ and the pair formation multiplicity $\kappa$.

More detailed calculations of the resulting spectrum and high-energy light curves in 3D will be presented in a future work. It is easy to obtain light curves since the source of the high-energy radiation is a narrow region at the origin of the equatorial current sheet. Obviously, for pair formation multiplicities $\kappa\lsim 10$ (or equivalently $\gamma$-ray efficiencies $L_\gamma/\dot{E}\gsim 10\%$), $\delta$ will be of the same order as $r_{\rm pc}$, and $\Delta$ will be of the same order or larger than $r_{\rm lc}$, and therefore, the global solution will differ significantly from the one presented in this work. That regime of pair formation multiplicities can only be investigated with numerical simulations \citep[e.g.][]{Ketal18}. We remind the reader, though, that \citet{CKK14} have already considered analytically the limit $\kappa=1$ and obtained $\gamma$-ray efficiencies $L_\gamma/\dot{E}\approx 40\%$. In that case, the equatorial dissipation zone extends all the way from the Y-point to infinity. Their solution was later confirmed via PIC simulations performed by \citet{CPPS15}. Finally, in the limit of no pair formation $\kappa\approx 0$, the solution is expected to degenerate to a `dead' pulsar with no electric currents.

%\section*{Acknowledgements}
%
%We thank 

%$$$$$$$$$$$$$$$$$$$$$$$$$$$$$$$

\bibliographystyle{mn2e}

\end{document}